\documentclass[11pt]{article}

\date{}  

\usepackage{graphicx}

\usepackage{url}
\usepackage{cite}
\usepackage{plain}

\usepackage{wrapfig}
\usepackage{graphics}
\usepackage[english]{babel}
\usepackage[leqno]{amsmath}
\usepackage{amssymb}
\usepackage{verbatim}
\usepackage[mathscr]{eucal}

\usepackage{amstext}
\usepackage{makeidx}
\usepackage{amsthm}

\usepackage{hyperref}
\hypersetup{colorlinks,%
citecolor=red,%
linkcolor=blue,%
urlcolor=black}

\makeindex

\newtheorem{Remark}{Remark} 
\newtheorem{Theorem}{Teorema} 

\newtheorem{proposition}{Proprieta'}
\newtheorem{definition}{Definizione}
\newtheorem{notation}{Nota}
\newtheorem{ex}{Esercizio} 
\newtheorem{esempio}{Esempio}

\newcommand{\beq}{\begin{equation}} 
\newcommand{\eeq}{\end{equation}}

\newcommand{\bex}{\begin{ex}} 
\newcommand{\eex}{\end{ex}} 

\newcommand{\bese}{\begin{esempio}} 
\newcommand{\eese}{\end{esempio}} 

\newcommand{\bpro}{\begin{proposition}} 
\newcommand{\epro}{\end{proposition}}

\newcommand{\bthe}{\begin{theorem}} 
\newcommand{\ethe}{\end{theorem}}

\newcommand{\bnote}{\begin{notation}} 
\newcommand{\enote}{\end{notation}}

\newcommand{\bdefi}{\begin{definition}} 
\newcommand{\edefi}{\end{definition}} 

\newcommand{\bc}{\begin{center}} 
\newcommand{\ec}{\end{center}}

\newcommand{\mail}[1]{\href{unina:#1}{\texttt{#1}}}

\usepackage{pdfsync}

\author{Monica De Angelis\thanks{Univ. of Naples  "Federico II", Dip. Mat. Appl. "R.Caccioppoli",
 \newline\mail{modeange@unina.it}}}

\title{Transport Phenomena in Excitable Systems: Existence of  Bounded Solutions  and Absorbing Sets}

\begin{document}
\maketitle

\abstract{In this paper, the transport phenomena of synaptic electric impulses are considered. The FitzHugh--Nagumo and FitzHugh--Rinzel models appear mathematically appropriate for evaluating these scientific issues. Moreover,  applications of such models arise in several biophysical phenomena in different fields such as, for instance, biology, medicine and electronics, where, by means of nanoscale memristor networks, scientists seek to  reproduce the behavior of biological synapses. The present article deals with the properties of the solutions of the FitzHugh--Rinzel system in an attempt to achieve, by means of a suitable ``energy function'', conditions ensuring the 
boundedness  and existence of absorbing sets in the  phase space. The results obtained depend on several  parameters characterizing the system, and, as an example, a concrete case is considered.}

\vspace{3mm} \textbf{Keywords} {Transport phenomena; FitzHugh--Rinzel model; Absorbing sets; Nonlinear dynamics; Biological neuron models}


\maketitle

\section{Introduction }
As is well 
known, transport phenomena are observable in a variety of scientific fields. The~one that will be taken into account here  is the transport of electric charges, a~phenomenon  that finds application in the most diverse areas.
In physiology, this event can be observed  in the {activities} of neurons, which are the fundamental units of the nervous system. Indeed, thanks to their peculiar physiological and chemical properties, neurons are able to receive, process, and~transmit electrical signals that, associated with ionic currents, cross the neuron's membrane. These electrical signals are called nerve impulses. The~difference in electrical charge existing between the inside and outside of the neuronal cell is called the membrane potential, while the variation in the membrane potential is called an action potential. Action potentials travel along the axon and are transmitted unchanged to other neurons in the form of electrical impulses. {This is only one of the multiple ways in which the complex functioning process of the so-called synapse happens. This process is well known in literature and is covered by an extensive bibliography} \cite{m1,j62,i,ks, 2000a}.
{In particular,} a reference point for these studies is the work of Hodgkin and Huxley (HH){,}  ref.~{\cite{h2}},
who developed a model of  propagation of the electrical signal along a squid axon. Their scheme consists of a system of four differential equations describing the dynamics of the membrane potential and the ionic current. However, as~the model was extended and applied to a wide variety of excitable cells, it became apparent that its  non-linearity and high dimensionality  did not allow it to perform a smooth analysis.    
Consequently, simpler models, such as the FitzHugh--Nagumo system (FHN) and the FitzHugh--Rinzel model (FHR), were introduced to allow the essential aspects of the dynamics of more complicated models to be~captured.

The bibliography in this regard is extensive, and  a very wide analysis exists (see, for instance, 
  refs.~{ \cite{gambino22,drw,dr13,r2,krs,mda18}} and references therein). Among~the many aspects, a~particularly interesting one concerns the equivalence that such mathematical models create between biological problems and the electrical transmission phenomena involved, such as in~the superconducting processes of Josephson junctions {\cite{nono,df213,dma18,scott,scott2}}. This suggests that the analysis of such models is reflected in both biological and superconducting transport~phenomena. 

Moreover, the~FHR system is able to describe the so-called bursting oscillations. This phenomenon occurs in a vast number of different cell types, and it is  characterized by an alternation between short bursts of oscillatory activity and periods of quiescence during which the membrane potential changes slowly {\cite{i, 2000a}}. Moreover, {bursting oscillation phenomena are} becoming increasingly important in many scientific fields in  light of its practical applications{.} As an example, some studies  on nanoscale memristor devices are directed to the restoration of synaptic connections by mimicking the behavior of  biological synapses and suggesting that electronic synapses could be introduced in the future to directly connect neurons {\cite{jnbbikem,llb}}. Moreover, electrical charge transport phenomena due to bursting oscillations have been observed in many nerve and endocrine cells, such as hippocampal and thalamic neurons, the~mammalian midbrain, and~the pancreas in {$\beta$-cells} 
 (see, for~example, ref.~\cite{bm} and references therein). This phenomenon also occurs in the cardiovascular system through the electrical activity of cardiac cells that excite the heart
to produce contractions of the ventricles and auricles~\cite{q17}. 

The previous  observations have aroused great interest and have led  to the conduction of, among~other things, an~analysis of FHR solutions. Indeed, several results regarding the existence of  exact solutions  have already been proved (see, for instance, ref.~{\cite{ZB,carillo,2020,moca21,lig}} and references therein), while more general analytical results can be found, for~example, in~{\cite{mda22,smoller,ccdf,R3}}.

In this article, the~FHR system is considered, and particular attention is paid to its solutions. {More specifically,  the~problem of the existence of attractors  and invariant sets is taken into account. Indeed, in qualitative analysis, this problem is of great importance; just think of its implications in issues concerning stability and a priori estimates.  Hence, } this paper aims to evaluate  the conditions of existence for both bounded solutions and absorbing sets as well as provide  examples of explicit~cases.

This paper is organized as follows: in Section~\ref{section2},  brief mathematical considerations on  the FHR physical  models are highlighted, while in Section~\ref{section3},  conditions guaranteeing solutions to boundedness are achieved, and an explicit example is {determined}. {Moreover, some numerical solutions are shown.} In Section~\ref{section4},  the existence of  absorbing sets {in the phase space} is  proved by giving an order of size according to  values stated to physical constants  characterizing the system. Finally, Section~\ref{section5} is devoted to a brief~comment.  

\section{Mathematical~Considerations }
\label{section2}
As  expected, terms in  FHR  reflect  peculiarities of  physical problems  and, as~a  general system, the~following~one

\begin{equation}
\label{12n}
\left \{
\begin{array}{lll}
	\displaystyle{\frac{\partial \,u }{\partial \,t }} =\,  D \,\frac{\partial^2 \,u }{\partial \,x^2 } -au \,\,+k u^2\, (\,a+1\,-u\,)
	\,-\, w\,\,+y\,\,  +I \,  \\
	\\
	\displaystyle{\frac{\partial \,w }{\partial \,t } }\, = \, \varepsilon (-\beta w +c +u)
	\\
	\\
	\displaystyle{\frac{\partial \,y }{\partial \,t } }\, = \,\delta (-u +h -dy)
\end{array}
\right.
\end{equation}

\noindent can be   taken into~consideration. 

Model (\ref{12n}) can be considered as a two time-scale slow-fast system with two fast
variables $  (u,w)$ and one slow variable $y$. However, if,~for instance, $\varepsilon =\delta,$ the system
can be considered as a two time-scale with one fast variable $ u $ and two slow variables
$  (w,y)$. Otherwise, if~$ \delta $  and $ \varepsilon $ have significant difference, it can also be considered as
a three-time-scale system with the fast variable $ u$, one intermediate variable and a
slow variable {\cite{xxcj}}.  {Moreover, $ ( D,a, I, \varepsilon ,\beta,c,d,  h,\delta,k  )$  indicate arbitrary real  constants,  and~if a   propagation phenomenon is considered, they assume specific meanings. As~an example, in~the biophysical activities of neurons,    $ a  $  represents the threshold constant  and  is  an excitability parameter}~\cite{GAR}{. Cases  with function $ a(x) $ or $ a $ depending on time have been  examined in~\cite{AD,fsdb}}.


For what concerns  the coefficient  $ D > 0, $  it    gives a  diffusion contribution, and~in synaptic studies, it depends on  the axial current in the axon. Indeed, it derives from the HH theory  where, if $ d $ represents the axon diameter and $ r_i$ is the resistivity, the~spatial variation in the potential $ V  $ gives the term $ (d/4 r_{i}) V_{xx} $,  from  which the term $ D\, u_{xx} $ is given~\cite{j62}.

{The parameter $ \varepsilon $ specifies the ratio between the time constants of the activator and the inhibitor~\cite{GAR}. Moreover, $I $ measures the amplitude of the external stimulus current and~it is  modulated by  variable $ y $  on a slower time scale~\cite{ks,r87}, while $ c$ and $ \beta $  can be  related to the number of channels of the cell membrane opened to  sodium and potassium ions, respectively~\cite{rr}. Moreover,  if~$  \beta \varepsilon  $ and $\delta d  $ are positive  constants, they  can  be considered  as  viscosity coefficients~\cite{R3}.}

With respect to the system  (\ref{12n}), {in some cases, the  constant $ k $ has been considered to  find explicit solutions (see, for instance,~\cite{krs}), and  } when  $ k=1, $ {denoting by}  
$ (u_0 ,w_0,y_0) $  the initial conditions, and {with}
\begin{equation}
\label{188}
\displaystyle F =u^2(a+1-u)+I- w_0 e^{- \varepsilon  \beta  t}+ y_0  e^{- \delta  d  t}- \frac{c}{\beta}( 1- e^{- \varepsilon \beta t} )+\frac{h}{d}( 1- e^{ -\delta d t} )
\end{equation}
being the non linear source function, the solution can be expressed by means of the integral equation given by: \cite{2020}
\begin{eqnarray}  \label{A14}
& \displaystyle \nonumber u (x,t)  =\int_\Re   \,H ( x-\xi, t)\,\, u_0 (\xi)\,\,d\xi \,
\\
\\
&\displaystyle \nonumber\,+\,\int ^t_0     d\tau \int_\Re   H ( x-\xi, t-\tau)\,\, F\,[\,\xi,\tau, u(\xi,\tau\,)\,]\,\, d\xi,
\end{eqnarray}

\noindent  where,  denoting by     $ J_1 (z) \,$    the  Bessel function of the first kind and order $\, 1,\,$   the  fundamental solution $ H(x,t) $ can be expressed by
\begin{equation} \label{A16}
\displaystyle H = H_1  - H_2,
\end{equation}

\noindent with

\noindent
\begin{equation} 
\nonumber \displaystyle    H_1(x,t)\,=\, \, \, \frac{e^{- \frac{x^2}{4\,D\, t}\,}}{2 \sqrt{\pi  D t } }\,\,\, e^{-\,a\,t}\, - \frac{1}{2} \,\,    \,\,\,\int^t_0  \frac{e^{- \frac{x^2}{4 \, D\, y}\,- a\,y}}{\sqrt{t-y}} \,\, \, \frac{\,\sqrt{\varepsilon} \,\, e^{-\beta \varepsilon \,(\, t \,-\,y\,)}}{\sqrt{\pi \, D \,}} J_1 (\,2 \,\sqrt{\,\varepsilon \,y\,(t-y)\,}\,\,)\,\,\} dy,
\end{equation}

\noindent and
\begin{equation} \label{H2}
\displaystyle H_2 =\int _0 ^t  H_1(x,y) \,\,e^{ -\delta d (t-y)} \sqrt{\frac{\delta y}{t-y}}   J_1( \,2 \,\sqrt{\,\delta \,y\,(t-y)\,}\,\,\, dy.
\end{equation}

For model (\ref{12n}), other properties have been proven, and in particular, by assuming $ \beta \varepsilon = \delta d, $ a class of explicit solutions has been~obtained in~\cite{moca21}.

In this paper,  our attention is devoted to  
the  following FHR model:

\noindent

{\protect \begin{equation}
	\label{22}
	\left \{
	\begin{array}{lll}
		\displaystyle{\frac{d \,u }{d \,t }} =\,   -au \,\,+ u^2\, (\,a+1\,-\frac{1}{k}\,u\,)
		\,-\, w\,\,+y\,\,  +I \,  \\
		\\
		\displaystyle{\frac{d \,w }{d \,t } }\, = \, \varepsilon (-\beta w +c +u)
		\\
		\\
		\displaystyle{\frac{d \,y }{d \,t } }\, = \,\delta (-u +h -dy)
	\end{array}
	\right.
\end{equation}}

\noindent where {all parameters are assumed to be real constants, and} if bursting phenomena are to be studied, the~physical variables $ (u,w,y) $ can be associated, respectively, to the transmembrane potential, the~recovery variable
and the slow-moving current in the dendrite. Specifically,~(\ref{22}) is characterized by two fast
variables $  (u,w),$ that show a relaxation oscillator in the phase plane where $ \varepsilon $ is a small parameter, and~by  the slow variable $y, $ whose pace is determined by the small parameter $ \delta. $ { In addition, as~usual, $I $ represents   } the amplitude of the external stimulus current.
A decrease in the value of the constant $ c $ causes longer intervals
between two bursts, while an~increase causes a shortening of the  intervals between bursts  and a change from  periodic bursting  to tonic~spiking.

System (\ref{22}) for $ a=-1 $ and $ k=3 $  turns~in

\noindent

{\protect\begin{equation}
	\label{FHR}
	\left \{
	\begin{array}{lll}
		\displaystyle{\frac{d \,u }{d \,t }} =\, u-u^3/3 + I_{ext} 
		\,-\, w\,\,+y\,\, \,  \\
		\\
		\displaystyle{\frac{d \,w }{d \,t } }\, = \, \varepsilon (-\beta w +c +u)
		\\
		\\
		\displaystyle{\frac{d \,y }{d \,t } }\, = \,\delta (-u +h -dy)
	\end{array}
	\right.
\end{equation}} 

\noindent which is often analysed in~literature.

\section{Conditions for Bounded~Solutions}
\label{section3}

Several methods have been developed to study the properties of solutions.  One that seems to be quite successful  is the method that takes into account the so-called ``energy function'' as, for~example, shown in~\cite{R3}.   Accordingly, the~following theorem can be proved: 

\begin{Theorem}  \label{th1}
Let us consider a FHR system (\ref{22}) and let us assume $ k>0. $ {Denoting} by 
\begin{equation}
	\label{ca11s}
	\left \{
	\begin{array}{lll}
		\displaystyle f(\beta, \varepsilon,a ) =\, \beta\,\varepsilon \, -\,\frac{|\varepsilon -1|}{2} \,-\, \,\, \frac{(1+a)^2\,k}{2}   \\
		\\
		\displaystyle g(\delta , d,a)= \,\delta \,d   \,- \frac{|1-\delta|}{2}\, -       \,\,   \frac{(1+a)^2\,k}{2},  
	\end{array}
	\right.
\end{equation}

\noindent 
if the system of constants is such that  the following conditions:

\begin{equation}
	\label{ca11ss}
	\left \{
	\begin{array}{lll}
		\displaystyle  f(\beta, \varepsilon,a ) >0   \\
		\\
		\displaystyle g(\delta , d,a)>0,  
	\end{array}
	\right.
\end{equation}

\noindent hold, then solutions of (\ref{22}) are~bounded. 

\end{Theorem}

\begin{proof}
Let us introduce  the  following energy function:

\begin{equation}
	{E} = \frac{1}{2} (u^2+w^2+y^2).
\end{equation}

Denoting by

\begin{equation}
	\eta = \beta \varepsilon; \qquad  \gamma = \delta d, 
\end{equation}

\noindent  system (\ref{22}) becomes

{\protect\begin{equation}
		\label{244}
		\left \{
		\begin{array}{lll}
			\displaystyle{\frac{d\,u }{d \,t }} =\, - a u + u^2(a+1) - \frac{1}{k} u^3 
			\,-\, w\,\,+y\,\, +I  \,  \quad \quad
			\\
			\\
			
			\displaystyle{\frac{d \,w }{d \,t } }\, = \,- \eta   w +\varepsilon\, c +\varepsilon \,u  \quad \quad 
			\\
			\\
			
			\displaystyle{\frac{d \,y }{d \,t } }\, = \,-\delta \,u +\delta\,h - \gamma \,y
		\end{array}
		\right.
\end{equation} }


Consequently, one has:

\noindent
\begin{eqnarray} 
	\nonumber & \displaystyle \frac{d {E}}{dt} = \displaystyle u \big[ \,- u a -u( \eta +\gamma) +u(\eta +\gamma)+ u^2(a+1) - \frac{1}{k} u^3 
	\,-\, w\,\,+y\,\, +I \big]+  
	\\
	\\
	\nonumber &  \displaystyle + w \big( - \eta   w +\varepsilon\, c +\varepsilon \,u\big) + y \big( -\delta \,u +\delta\,h - \gamma \,y \big)
\end{eqnarray}

\noindent 
or rather
\begin{eqnarray} \label{282ffv}
	\nonumber &  \displaystyle  \frac{d {E}}{dt} = - u^2  (  \eta +\gamma) +  u^2 (-a+ \eta +\gamma) + u^3 (a+1) - \frac{u^4}{k}
	\\
	\\
	\nonumber &  \displaystyle + u( \,-\, w\,\,+y\,\, +I)+ w ( - \eta   w +\varepsilon\, c +\varepsilon \,u )+ y \,( -\delta \,u +\delta\,h - \gamma \,y ).
\end{eqnarray}

Considering that it results in
\begin{eqnarray}
	\nonumber &  \displaystyle u^2 \,(- a + \eta +\gamma )\leq \,\, ( -a + \eta +\gamma )^2 \,\,\frac{k}{2}  +  \,\,\frac{u^4}{2\,k}  \\ \\
	\nonumber &  \displaystyle 
	u^3 (a+1) \leq  \dfrac{u^4 }{2\,k} \,+ u^2( a+1)^2 \, \frac{k}{2},
\end{eqnarray}

\noindent  
from (\ref{282ffv}), one deduces:

\noindent
\begin{eqnarray} \label{282ccvq}
	\nonumber & \displaystyle \frac{d {E}}{dt} \leq
	- \,u^2  \bigg[   \eta +\gamma -\frac{(a+1)^2}{2}\,k\,\bigg]-\eta w^2 -\gamma\, y^2\, + (\varepsilon-1) \,w \,u  +(1-\delta) \,y\, u   
	\\
	\\
	\nonumber &  \displaystyle +uI \,   +w \varepsilon\, c +   \, +\delta\,h y + \frac{1}{2} (- a + \eta +\gamma )^2 k.  
\end{eqnarray}

\noindent

Now, introducing an arbitrary constant  $ \varepsilon_1 \geq 0,  $  let

\begin{equation}  \label{A}
	A= (a+1)^2\,k \,+ 2\varepsilon_1.
\end{equation}

Since it results in

\begin{equation}
	\label{rrvvvq}
	\left \{
	\begin{array}{lll}
		\displaystyle  (\varepsilon -1)wu \leq  \frac{|\varepsilon -1|}{2} (w^2+u^2); \qquad \delta h y \leq \frac{h ^2 \delta ^2}{2 A}+ \frac{A y^2}{2} 
		\\
		\\
		\displaystyle   Iu \leq \frac{I^2}{ 2 A} + \frac{A u^2}{2}; \qquad   w \varepsilon\, c \leq \frac{\varepsilon ^2 c^2}{2 A} + \frac{A w^2}{2};
		\\
		\\
		\displaystyle  (1-\delta) uy \leq \frac{|1-\delta|}{2} (u^2+y^2),    
	\end{array}
	\right.
\end{equation}

\noindent    from (\ref{282ccvq}), one obtains:
\begin{eqnarray} \label{28ccvvq}
	\nonumber & \displaystyle \frac{d {E}}{dt} \leq
	-  u^2 \bigg[    \eta +\gamma -\frac{(a+1)^2}{2} k - \frac{A}{2}  - \frac{|\varepsilon -1|}{2} - \frac{|1-\delta|}{2} \bigg]
	\\
	\\   \nonumber
	&  \displaystyle -  w^2  \bigg( \eta - \frac{A }{2} -\frac{|\varepsilon -1|}{2}\bigg)\,-  y^2  \bigg(\gamma -\frac{|1-\delta|}{2}-\frac{A }{2}\, \bigg) 
	\\\nonumber
	\\ \nonumber\\ \nonumber
	\nonumber &  \displaystyle +\frac{1}{2A} \big(I^2+ \,   h ^2 \delta ^2+\varepsilon ^2 c^2\big)   + \frac{1}{2} (- a + \eta +\gamma )^2  k. 
\end{eqnarray}

Therefore, for (\ref{A}), it results in

\begin{eqnarray} \label{28ccvvq}
	\nonumber & \displaystyle \frac{d {E}}{dt} \leq
	-  u^2 \bigg[    \eta +\gamma -(a+1)^2\, k - \varepsilon_1  - \frac{|\varepsilon -1|}{2} - \frac{|1-\delta|}{2} \bigg]
	\\
	\\   \nonumber
	&  \displaystyle -  w^2  \bigg( \eta - \frac{(a+1)^2\,k}{2} \,-\varepsilon_1 -\frac{|\varepsilon -1|}{2}\bigg)\,-  y^2  \bigg(\gamma -\frac{|1-\delta|}{2}- \frac{(a+1)^2\,k}{2} \,-\varepsilon_1\, \bigg) 
	\\\nonumber
	\\ \nonumber\\ \nonumber
	\nonumber &  \displaystyle +\frac{I^2+ \,   h ^2 \delta ^2+\varepsilon ^2 c^2}{4\varepsilon_1+ 2\,k \,(a+1)^2}   + \frac{\,(- a + \eta +\gamma )^2  \,k}{2}.\, 
\end{eqnarray}

According to hypotheses (\ref{ca11s}) and (\ref{ca11ss}), it is possible to fix $ \varepsilon_1 \geq 0 $ such that :

\begin{equation}
	\label{cassio}
	\left \{
	\begin{array}{lll}
		\displaystyle f(\beta, \varepsilon,a ) =\, \beta\,\varepsilon \, -\,\frac{|\varepsilon -1|}{2} \,-\, \,\, \frac{(1+a)^2\,k}{2}  >\varepsilon_1  \\
		\\
		\displaystyle g(\delta , d,a)= \,\delta \,d   \,- \frac{|1-\delta|}{2}\, -       \,\,   \frac{(1+a)^2\,k}{2} >\varepsilon _1.  
	\end{array}
	\right.
\end{equation}

As a consequence, when  conditions (\ref{cassio}) hold, it also results in

\begin{equation}
	\displaystyle\eta  + \gamma  - k\,(a+1)^2 \,- \varepsilon_1  -\frac{|\varepsilon -1|}{2} - \frac{|1-\delta|}{2} > \varepsilon_1.
\end{equation}

Thus, denoting by
\begin{equation}
	\label{rrvv12q}
	\left \{
	\begin{array}{lll}
		\displaystyle  B= \displaystyle \eta -\frac{|\varepsilon -1|}{2}  - \frac{(a+1)^2\,k}{2} \,-\varepsilon_1 >0 
		\\
		\\
		\displaystyle  B_1= \gamma -\frac{|1-\delta|}{2}- \frac{(a+1)^2\,k}{2} \,\,\,-\varepsilon_1 > 0
	\end{array}
	\right.
\end{equation}

\noindent and

\begin{equation}
	\label{rrvv12q1}
	\left \{
	\begin{array}{lll}
		\displaystyle   \,C= 2\, \min\,\{B,B_1\}
		\\
		\\
		\displaystyle  C_1 = \frac{I^2+ \,   h ^2 \delta ^2+\varepsilon ^2 c^2}{4\varepsilon_1+ 2\,k\,(a+1)^2}   + \frac{\,(- a + \eta +\gamma )^2  \,k}{2}\,  
	\end{array}
	\right.
\end{equation}

From (\ref{28ccvvq}), one gets:

\begin{equation}
	\frac{d {E}}{dt} \leq -C {E} +C_1.
\end{equation}

Consequently, it follows that:

\begin{equation} \label{103}
	{E} \leq  \frac{C_1}{C}   (1-e^{-Ct})   + {E}_0 e^{-Ct},
\end{equation}

\noindent  from which, $ \forall t \geq 0,  $ one obtains

\begin{equation} \label{E}
	{E} \leq  {E}_0 +\frac{C_1}{C}. 
\end{equation}
\end{proof}

\begin{Remark}
Naturally, a constant $ \varepsilon _1 $ for which the (\ref{cassio}) are worth has to be determined according to the values taken by all variables of the (\ref{ca11s}). 
By proving a possible application,  it will  show how  variable $a$ and $\varepsilon _1$ are related to each~other.  
\end{Remark} 

There are many examples in  literature of numerical values assigned to the FHR system (see, for instance, ref. {\cite{R3}}). Here, just for instance, the~following set is considered:

\begin{equation}  \label{ee}
\left \{
\begin{array}{lll}
	\displaystyle I= 0.3125\quad \varepsilon= 0.8 \quad  c= 0.2 \quad h=-0.775  
	\\
	\\
	\displaystyle  \beta = 0.126 \quad \delta= 0.5 \quad d=1; \quad {k}=3.
\end{array}
\right.
\end{equation}

To prove the existence of  constant $ \varepsilon_1 \geq 0  $ such that:

\begin{equation}
\label{rrvv1qff}
\left \{
\begin{array}{lll}
	\displaystyle \eta -\frac{|\varepsilon -1|}{2}  - \frac{(a+1)^2\,k}{2} \, >\varepsilon_1
	\\
	\\
	\displaystyle  \gamma -\frac{|1-\delta|}{2}- \frac{(a+1)^2\,k}{2} \,\,> \varepsilon_1,
\end{array}
\right.
\end{equation} 

\noindent let us require, firstly,  that  conditions

\begin{equation}
\label{uuu}
\left \{
\begin{array}{lll}
	\displaystyle \eta -\frac{|\varepsilon -1|}{2}  - \frac{(a+1)^2\,k}{2} \, >0          \\
	\\
	\displaystyle  \gamma -\frac{|1-\delta|}{2}- \frac{(a+1)^2\,k}{2} \,\,>0
\end{array}
\right.
\end{equation}

\noindent  are satisfied.  
For this, since the minimum value between  $ (B,B_1) $  is independent from variables $ a, \varepsilon_1 $ and  since:

\begin{equation}
\left \{
\begin{array}{lll}
	\displaystyle  \eta = \beta\, \varepsilon =  0.1008;  \qquad   \gamma= \delta \,d = 0.5 
	\\
	\\   
	\displaystyle  \frac{|1-\varepsilon|}{2} = 0.1 \qquad   \displaystyle  \frac{|1-\delta|}{2} = 0.25 , 
\end{array}
\right.
\end{equation}

\noindent it is deduced that

\begin{equation} \label{aaa}
C= 2\bigg[ 0.0008- \frac{3\,(a+1)^2\,}{2} - \varepsilon_1 \,\bigg].
\end{equation}

Consequently,  it will be sufficient to choose variable $ a $ such that   $\,\, \displaystyle -1- \sqrt{3}/75  <a < \sqrt{3}/75 -1, $ (i.e, in an approximate form:   $\,\, \displaystyle -1.023094011 <a <-0.9769059892 $) to satisfy~(\ref{uuu}).

Hence, if, for~instance, $\displaystyle  a= -0.98,  $ in order to prove inequalities in  (\ref{rrvv1qff}), a constant $ \varepsilon_1  $ can be chosen in the {interval} 
 $ [0, 0.0002] $. 

This shows  that a concrete case with bounded solutions exists. Naturally, the~range of variation for variable $ a $ ensures that, even just by the system of parameters set in (\ref{ee}), several other explicit cases can be~obtained.  

\begin{figure}
\includegraphics[width=15cm, height=6cm]{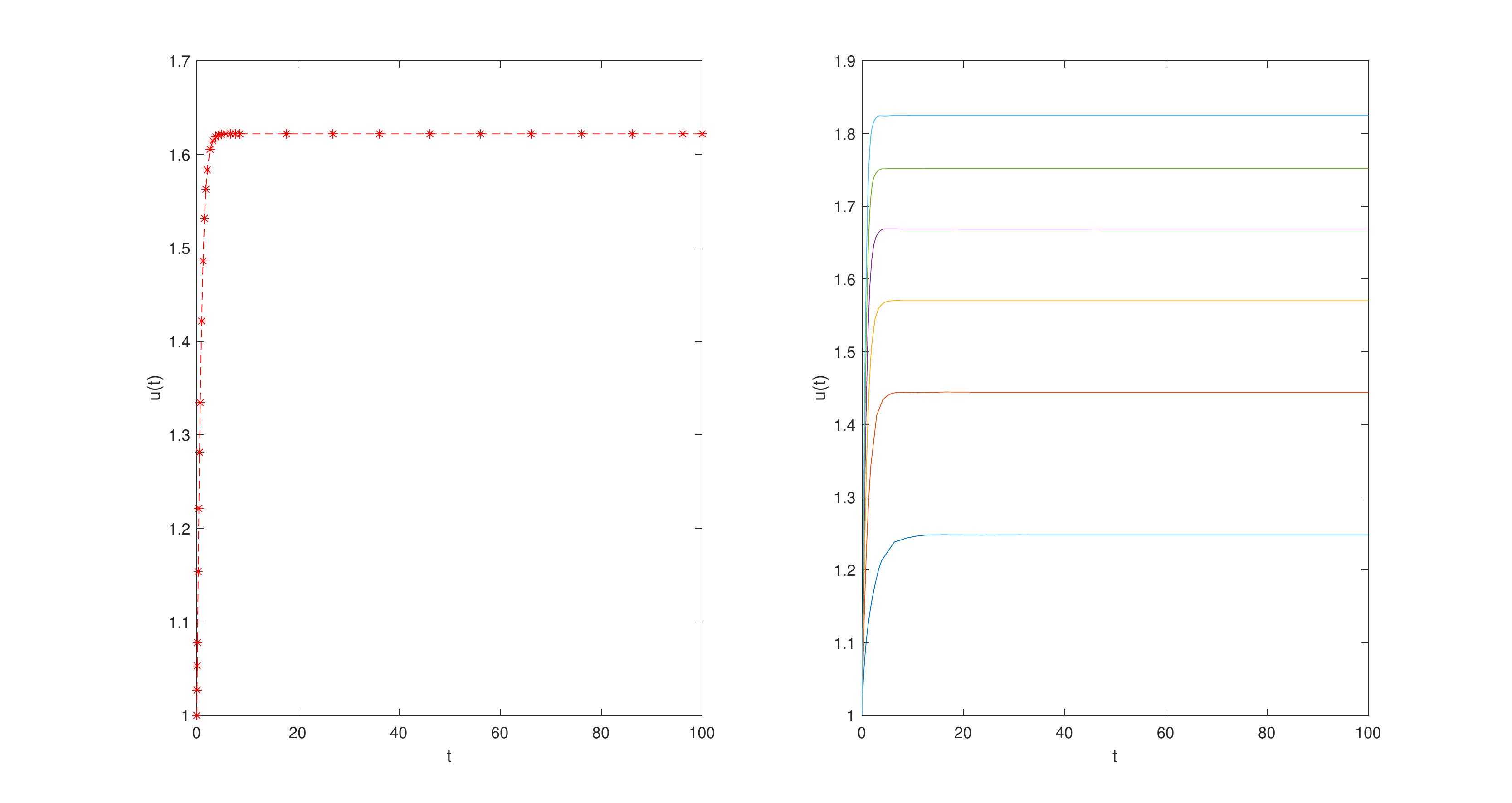} \hfil
\caption{On the left: 
	the bounded solution \emph{u}(\emph{t}) when  $ \varepsilon \beta=3.6, $  $u_0=1  $, $C_1= -0.2, $ $ C_2= 0.2. $ On the right: the bounded functions $u(t)$ as $-0.6\leq C_1 \leq 0.2 $.
}
\label{etichetta}
\end{figure}

\begin{Remark}
Let us consider the FHR model expressed in (\ref{FHR}).  In~the hypothesis that $ \varepsilon \beta= \delta d  $ and $ \varepsilon = -\delta, $ it is possible to prove that (\ref{FHR}) admits the following first integral:

\begin{equation} \label{36}
	\frac{du}{dt}+\frac{1}{3}u^3- u = C_1 + C_2 e^{- \beta \varepsilon \,t},
\end{equation}

\noindent where $ C_1 $ is a constant depending on $ (I, c, h, \beta) $ and  $ C_2\neq 0 $ is an arbitrary constant.

By means of (\ref{36}), the~solution has been obtained through the Matlab solver $  ode15s$, and in Figure \ref{etichetta}
graphs show how the  solution $ u(t) $ remains~bounded. 
\end{Remark}

\section{Existence of Absorbing~Sets }

\label{section4}

Let us consider bounded solutions. As~previously observed, the~existence  of bounded solutions also depends on the choice of the variable $ a $. Consequently, once it is determined to be an absorbing set, it will be possible to give it  an  order of magnitude in agreement with the values of the parameters of the system,  and~according to the choice of   values of $ a $  and $ \varepsilon_1. $

In searching for  an absorbing set, it is necessary to make sure that it is  both invariant  and an attractor.  For~this, and also taking into account~\cite{R3}, the~following theorem is proved:

\begin{Theorem}

In the hypotheses  of theorem  (\ref{th1}), indicating by $ {K_0} $ a positive constant, let us assume $ \displaystyle  E_0  < {K_0}. $ Besides, denoting by   $ D_R $ the sphere  of the phase space  of center the origin and radius  less than  $R,$ let us assume

\begin{equation}
	R= \, \sqrt{2}\,\,\sqrt{\,\frac{ \,C_1}{C}+  {K_0}}\,.
\end{equation}


Then, $ D_R  $ proves to be an absorbing  set for the system (\ref{22}).

\end{Theorem}

\begin{proof}
Theypothesis on $ E_0  $   and relation (\ref{E}) assure that:

\begin{equation} \label{ssE}
	{E} \leq  {E}_0 +\frac{C_1}{C}< {K_0}+ \frac{C_1}{C}, 
\end{equation}

\noindent  and consequently,  $ D_R $ is~invariant.

Moreover,   denoting by  $ B $ a bounded region of the phase space,  let us~define

\begin{center}
	$ \displaystyle \tilde E_0 =\max_ B  E_0,  $
\end{center}

\noindent and let us consider bounded solutions of system (\ref{22}) whose initial  data $(u_0,w_0,y_0)$  belong to~$ \displaystyle \tilde E_0.  $

Because of  (\ref{103}), it results that

\begin{equation}  \label{v} 
	\displaystyle   {E}   \leq  e^{-Ct} \,\bigg|{\tilde E}_0- \frac{C_1}{C} \bigg| + \frac{C_1}{C}   \qquad \forall t\geq 0
\end{equation}   

\noindent and hence, there exists a positive \emph{T} such that $ \forall t>\,T  $  one~has

\begin{equation}  
	\displaystyle   {E (t)}   \leq  \frac{C_1}{C}. 
\end{equation}

On the other hand, formula (\ref{v}) leads us to consider   the positive   instant    $ \tau  $ given~by

\begin{equation}  \label{r}
	r^2 = \frac{C_1}{C} + \bigg|\tilde E_0 -\frac{C_1}{C}\bigg| e^{-C\tau} 
\end{equation}

\noindent  which means:

\begin{equation}  \label{rrr}
	\tau = \frac{1}{C} \log \,\,\displaystyle \frac{\big|\tilde E_0 -\frac{C_1}{C}\big|}{\big|r^2-\frac{C_1}{C}\big|}, 
\end{equation}

\noindent  and it will be sufficient to assume $ r^2 = R^2/2 $ to get, for~all  $ t>\tau:   $

\begin{equation}  
	\displaystyle  {E}   \leq  e^{-C\tau} \,\bigg|{\tilde E}_0- \frac{C_1}{C} \bigg| + \frac{C_1 }{C}  = \frac{R^2}{2}.
\end{equation}
\end{proof}

\begin{Remark}
{Since} 
the size of the absorbing set $ \displaystyle D_R $ also depends on the radius  
$ \displaystyle R = \,\sqrt{\frac{2\,C_1}{C}}, $ it seems interesting to give an order of amplitude of $ \displaystyle R $ as a function of both  constants $ \displaystyle( I, \varepsilon, \beta, c, d, h, \delta, k) $ 
and of the values of $ a $ and $ \varepsilon_1 $ that can be chosen~accordingly.    
\end{Remark}

For this, as~an example, let us assume the set of values (\ref{ee}).

{Since   $-1.023094010768 <a <-0.9769059892324 $, choosing, for~instance,}\linebreak \mbox{$a~=~-0.9769059892 $} and using 10 digits of precision,   from~(\ref{aaa}),  we deduce that:

\begin{equation}
C\,=  2\,(0.00719999999998 - \varepsilon _1 \, )
\end{equation}

\noindent
and  assuming, for~instant,   $ \varepsilon _1  = 0.007199999997,   $  one {has:} 

\begin{equation}
C\,= 1.999999125 \times10^{-12}\sim
2 \times 10^{-12}
\end{equation}
then for what concerns $ C_1,  $  for $ a= - 1.0230940107 $ and $ \varepsilon_1=0,$ we obtain that
\begin{equation}
C_1\thickapprox 85.7089051.    
\end{equation}

Consequently:
\begin{equation}
\sqrt{\frac{ 2\,C_1}{C}} \thickapprox 9.2579 \times 10^6.
\end{equation}

\section{{Results and~Implications}}
\label{section5}

The paper deals with an analysis of the ternary nonlinear dynamical FitzHugh--Rinzel system. Properties of  solutions are investigated  and,  by~means of a suitable “energy function”, conditions that ensure the boundedness and existence of absorbing sets in the energy phase space are given. Moreover, according to the parameters characterizing the system, an~example has been considered showing that,  even with only one choice of parameters, the~variable $ a $ allows us to obtain several classes of bounded solutions. {Moreover, by choosing the value of parameters in accordance with the assumptions of Theorem 1,  by~means of a first integral of the FHR system {(7),} 
  some  graphs of numerical solutions have been obtained, showing bounded solutions as   predicted by  theory.}

{The present results will be quite useful when the analysis is turned to a different physical case. Indeed, the~constant $ a $ introduced in the FHR model (\ref{22}) generalizes the FHR system (\ref{FHR}), and~the results obtained do not directly involve the limits on $ a, $ thus suggesting the possibility of generalizing the results. 
}

{Moreover,} {the} scientific methodology employed can be applied in multiple scientific fields given the equivalence that such a mathematical model creates between biological problems and electrical transmission phenomena, such as in~the superconducting processes of Josephson junctions. {In this type of  junction,   a~third-order partial differential  equation, similar to the one introduced in~\cite{j62} and  extensively discussed in~\cite{nono,df213,dma18,scott,scott2}, describes the motion of squids in superconductivity.
}This suggests that the analysis of such models is reflected in a vast number of realistic mathematical models.
{Finally, it is important to underline that  the existence of bounded solutions and absorbing sets paves } the way for further research both  into stability {problems} and  in the Hopf bifurcations {that are so essential for the study of bursting oscillations}.

\textbf{Acknowledgments}{The  paper has been performed under the auspices of G.N.F.M. of~INdAM. {The author is grateful to the three anonymous reviewers for their comments and suggestions.} 
}

\end{document}